\documentclass[%
 reprint,
 amsmath,amssymb,
 aps,
 superscriptaddress,
 pre,
 twocolumn,
 titling
]{revtex4}

\usepackage{graphicx}
\usepackage{dcolumn}
\usepackage{bm}
\usepackage{braket}
\usepackage{mathrsfs}

\begin{document}

\preprint{APS/123-QED}

\title{Shortcut engineering of active matter: run-and-tumble particles}

\author{Adam G. Frim}%
\affiliation{%
 Department of Physics, University of California, Berkeley, Berkeley, California, 94720
}%
\author{Michael R. DeWeese}
\affiliation{%
 Department of Physics, University of California, Berkeley, Berkeley, California, 94720
}%
\affiliation{%
Redwood Center For Theoretical Neuroscience,  University of California, Berkeley, Berkeley, California, 94720
}
\affiliation{%
Helen Wills Neuroscience Institute,University of California, Berkeley, Berkeley, California, 94720
}%

\date{\today}

\begin{abstract}
Shortcut engineering consists of a class of approaches to rapidly manipulate physical systems by means of specially designed external controls. In this Letter, we apply these approaches to run-and-tumble particles, which are designed to mimic the chemotactic behavior of bacteria and therefore exhibit complex dynamics due to their self-propulsion and random reorientation, making them difficult to control. Following a recent successful application to active Brownian particles, we find a general solution for the rapid control of 1D run-and-tumble particles in a harmonic potential. We demonstrate the effectiveness of our approach using numerical simulations and show that it can lead to a significant speedup compared to simple quenched protocols. Our results extend shortcut engineering to a wider class of active systems and demonstrate that it is a promising tool for controlling the dynamics of active matter, which has implications for a wide range of applications in fields such as materials science and biophysics.
\end{abstract}

\maketitle


\indent \textit{Introduction.}--
Over the past several decades, active matter has become a major focus of study in nonequilibrium statistical mechanics. Inspired by interesting, novel observations in biological settings, such as self-propulsion \cite{rothschild_Nature_1963,berg_nature_1973,Wadhams_NatRevMCB_2004} and flocking \cite{viscek_PRL_1995,Toner_PRL_1995}, active matter consists of systems in which individual particles are able to dissipate energy from internal or other nonthermal energy reservoirs. Allowing for such internal energy dissipation and entropy production makes such systems inherently out of equilibrium. Active matter displays numerous unique, and at times universal, phenomena, such as pattern formation \cite{Farrell_PRL_2012}, motility induced phase separation \cite{Fily_PRL_2012,cates_AnnRev_2015,Digregorio_PRL_2018}, velocity alignment \cite{Caprini_PRL_2020}, and accumulation at confining walls \cite{Berke_PRL_2008,Fily_SoftMat_2014,Yang_SoftMat_2014,Solon_NatPhys_2015,Takatori_NatCom_2016}. Recently, the problem of controlling and manipulating biological as well as synthetic active matter systems has gained significant interest as a means for novel engineering at the nanoscale \cite{Needleman_NatRevMat_2017,Ross_Nature_2019}, with numerous potential applications, such active matter engines \cite{Sokolov_PNAS_2009,Ekeh_PRE_2020,Pietzonka_PRX_2019,Speck_PRE_2022,Fodor_EPL_2021}, designed self-assembled structures \cite{Schwarz-Linek_PNAS_2012,Sanchez_Nature_2012,Stenhammar_SciAdv_2016,Mallory_AnnRevPChem_2018,Scacchi_PRR_2021}, and programmable active matter devices \cite{Lum_PNAS_2016,Fu_2022,Gardi_NatCom_2022}.

Simultaneously, significant progress has been made in the field of shortcut engineering \cite{Berry_JPA_2009,Emmanouilidou_PRL_2010,PRL_Chen_2010,del-Campo_PRL_2013,Torrontegui_2013_chapter,Deffner_PRX_2014,2016_NP_Martinez_ESE,del_Campo_NJP_2019,2019_RMP_Guery,Guery-Odelin_RPP_2023}. In particular, for a system whose steady state $\rho_{\text{ss}}$ is specified by a set of external control parameters $\mathbf{\Lambda}$,  shortcut engineering is a set of approaches to modulate these  external controls such that a system in an initial steady state $\rho_0 = \rho_{\text{ss}}(\mathbf{\Lambda}_0)$ arrives at a target final steady state $\rho_f = \rho_{\text{ss}}(\mathbf{\Lambda}_f) $ in a finite time $\tau$. The protocol duration $\tau$ is typically chosen to be shorter than the natural relaxation timescale of the system $\tau_r$, such that such a protocol \textit{shortcuts} the system's relaxation. These finite-time approaches first gained attention in closed quantum systems, in which such approaches are known as shortcuts to adiabaticity as they shortcut the adiabatic theorem of quantum mechanics \cite{Berry_JPA_2009,Emmanouilidou_PRL_2010,PRL_Chen_2010}. More recently, shortcut engineering has been applied to significantly more complicated systems in both quantum and classical systems, both in the case of deterministic \cite{2017_PRE_Jarzynski,2017_NJP_Patra} and stochastic dynamics \cite{2016_NP_Martinez_ESE,2017_PRE_Geng,2018_NJP_Chupeau,Dann_PRL_2019,2020_SciP_Salambo,2021_NP_Iram,2021_PRE_Frim}, and even to driving rapidly between two nonequilibrium steady states for out-of-equilibrium systems \cite{Baldassari_PRL_2020,baldovin_2022} We emphasize that such shortcut protocols typically require the system to be out-of-steady-state with respect to current external controls at intermediate times, but the system arrives at a final steady state corresponding to  $\mathbf{\Lambda}(\tau)$ at the end of the protocol, such that after control modulation is concluded, the system remains in this new 
steady state.

In this Letter, we apply shortcut engineering to the paradigmatic active matter system of run-and-tumble particles (RTPs). An RTP is an overdamped colloidal
particle with a given self-propulsion velocity that stochastically and instantaneously redirects to a random direction at a fixed rate, similar to the chemotactic behavior of bacteria \cite{rothschild_Nature_1963,Wadhams_NatRevMCB_2004,Tailleur_PRL_2008} Specifically, inspired by recent work on another popular active matter system, active Brownian particles (ABPs) \cite{baldovin_2022}, we find a class of shortcut solutions that allow for rapid transitions between two steady-state distributions at different self-propulsion velocities for a system of non-interacting RTPs in a one-dimensional harmonic trap. Our results are experimentally testable with
existing techniques \cite{Takatori_NatCom_2016,Dauchot_PRL_2019,Schmidt_NatComm_2021}. In addition, we derive
the optimal protocols that minimize the average external work for such a modulation. Our findings could influence
the design of near-term experiments and provide insights needed to faithfully and rapidly control, manipulate, and engineer active systems.

\begin{figure*}
    \centering
    \includegraphics[width=\linewidth]{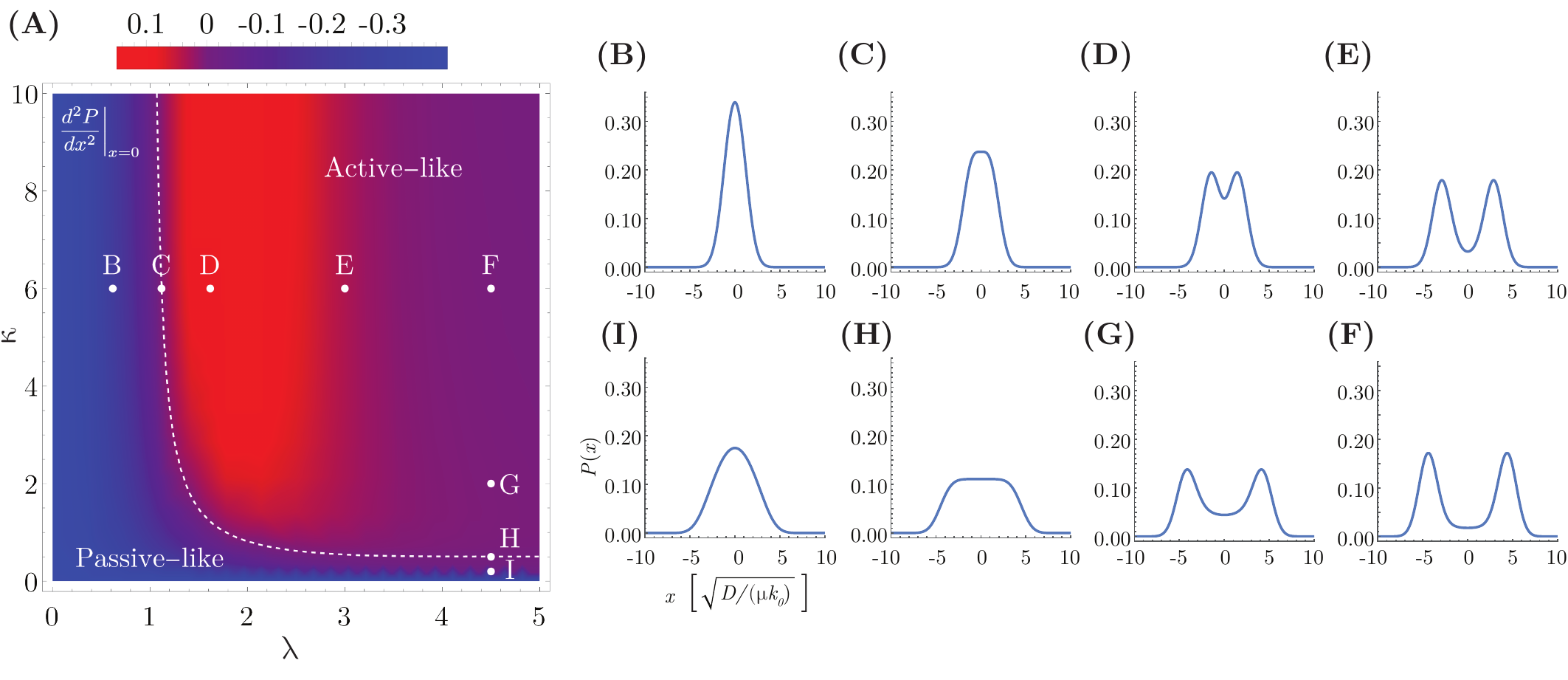}
    \caption{Steady-state distribution of run-and-tumble particles for a variety of parameter values. (A) The second derivative of the steady-state distribution at the trap center as calculated from the first 100 terms of Eq~\eqref{eq:steady-state}: a local maximum at the trap center denotes a passive-like state whereas a local minimum denotes an active-like state. The transition between these two regimes is denoted by the dotted line. (B)-(I) The steady-state distribution from the first 100 terms in Eq~\eqref{eq:steady-state} for the parameter values denoted in (A). All axes for (B)-(I) are the same as denoted in (I). Note that  behavior progresses from left to right along both rows, \textit{i.e.} (B) and (I) are both passive-like, (C) and (H) are both marginal, and (D)-(G) are all active-like.}
    \label{fig:fig1}
\end{figure*}

\indent \textit{Run-and-tumble dynamics in 1D.}--
RTP particles undergo dynamics given by

\begin{equation}
\label{eq:RTP_dyn_1}
\dot{x} = -\mu \partial_x U +v_0 \sigma(t) + \eta(t),
\end{equation}
where $\mu$ is the particle's mobility coefficient, $U(x)$ is a given confining potential profile, $v_0$ is the RTP self-propulsion velocity, $\sigma$ is telegraph noise which assumes the values $\sigma \in \{+1,-1\}$ and transitions in a Poisson process with given average rate $\gamma/2$. Its autocorrelation function is $\braket{\sigma(t')\sigma(t)} = e^{-\gamma |t-t'|}$. $\eta$ is Gaussian white noise due to a thermal heat bath with mean zero and $\braket{\eta(t)\eta(t')} = 2D\delta (t-t')$ for diffusion constant $D$. We specialize to the case of a harmonic trap where $U(x) = \frac{1}{2}  k x^2$ such that Eq.~\eqref{eq:RTP_dyn_1} becomes
\begin{equation}
\label{eq:RTP_dyn_2}
\dot{x} = -\mu kx +v_0 \sigma(t) + \eta(t),
\end{equation}
Following standard methods, we may write an equivalent master equation for the distribution of the particles $P(x,t)$:

\begin{subequations}
\label{eq:master_eq}
\begin{align}
&\partial_t P_+ = -\partial_x[(-\mu kx+v_0)P_+] + D\partial_x^2 P_+ -\frac{\gamma}{2}(P_+-P_-) \\
&\partial_t P_- = -\partial_x[(-\mu kx-v_0)P_-] + D\partial_x^2 P_- -\frac{\gamma}{2}(P_--P_+) ,
\end{align}
\end{subequations}
where $P_+ (P_-)$ refers to the probability distribution of the of the right- (left-) moving particles. The final term in each equation refers to the gain and loss between the different particle species whereas the first terms characterize intraspecies time evolution for each particle type. 

In steady-state, the distribution must satisfy
\begin{equation}
\label{eq:steady_state_master_eq}
0 = -\partial_x[(-\mu kx\pm v_0)P_+] + D\partial_x^2 P_+ \pm \frac{\gamma}{2}(P_--P_+) .
\end{equation}
Recent endeavors in finding exact solutions to active matter steady-state distributions \cite{Wagner_JSM_2017,Malakar_JSM_2018,Basu_PRE_2019,Malakar_PRE_2020,Chaudhuri_JSM_2021,Caraglio_PRL_2022} have led to the run-and-tumble and distribution in both thermal \cite{Dhar_PRE_2019} and athermal ($D=0$) cases \cite{Garcia-Millan_JSM_2021} 
\begin{widetext}
\begin{equation}
\label{eq:steady-state}
    P_{\pm}(x)^* =  \sum_{n=0}^\infty \left(\frac{v_0}{\sqrt{ \mu k D}} \right)^n \frac{1}{2} \sqrt{\frac{\mu k}{2\pi D}}e^{-\frac{\mu k x^2}{2D}}  (\pm 1)^n C_n(\frac{\gamma}{\mu k})  \textit{\text{He}}_n\left(\sqrt{\frac{\mu k}{D}} x\right),
\end{equation}
\end{widetext}
where $C_0(x) \equiv 1$ and
\begin{equation}
    C_n(x) = 
    \prod_{j=1}^n \frac{1}{j+\frac{1}{2}(1+(-1)^{n+1})x}, \ \ \ \  n>0.
\end{equation}
$\textit{\text{He}}_n$ is the $n$th Hermite polynomial under the ``probabilist's" convention:
\[\textit{\text{He}}_n(x) \equiv  e^{x^2/2} \frac{d^n}{dx^n} e^{-x^2/2},\]
and $P_{\text{tot}}(x) = P_+(x) + P_-(x)$. Note that this set of dynamics is characterized by two dimensionless quantities: the dimensionless stiffness $\kappa = \mu k/\gamma$  and the dimensionless velocity $\lambda  = v_0 /\sqrt{\mu k D}$. $\kappa$ encodes the competition of the characteristic timescales of particle transmutation $1/\gamma$ and Brownian particle relaxation $1/(\mu k )$. $\lambda$, in contrast, represents the competition between thermally driven diffusion and the directed motion due to the active propulsion. 

The steady-state behavior of the system is entirely set by these two parameters and one observes both passive-like states, encoded by the particles' tendency to accumulate at the bottom of the potential, and active-like states, encoded by the particles' tendency to accumulate at turning points in the potential. See Fig.~\ref{fig:fig1} for a steady-state phase diagram, as measured by the sign of $\partial_x^2 P_{\text{tot}}|_{x=0}$ calculated from the first 100 terms of Eq.~\eqref{eq:steady-state}. One can observe that, for $\lambda \rightarrow \infty$ (as $D\rightarrow 0$), the active-passive transition occurs exactly at $\kappa = 1/2$ whereas higher values of $\kappa$ are required for smaller $\lambda$ as the diffusive forces begin to ``wash out" the accumulation of particles at the boundaries.

\indent \textit{Shortcut solution.}--
We now wish to implement a shortcut strategy for RTP dynamics. Namely, we assume the distribution is initialized in the steady-state corresponding to the control parameters at time $t=0$. We wish to carry out some manipulation of the control parameters over some finite time horizon $0\leq t\leq \tau$ so as to arrive at a final state corresponding to the steady state for the given control parameters at finite time $\tau$. For certain parameter regimes, such a manipulation can result in rapid transitions from one steady state to another over timescales orders of magnitudes shorter than natural relaxation following a control parameter quench. Inspired by a recent work on ABPs \cite{baldovin_2022}, we will focus on the case of shortcutting between two different self-propulsion velocities, $v_0(0) = v_{00}, v_0(\tau) = v_{0f}$. We will attempt an inverse engineering approach, choosing an ansatz for the exact evolution of the distribution and arriving at necessary conditions on controllable external parameters to realize such a distribution. We consider the same ansatz as \cite{baldovin_2022}:
\begin{equation}
\label{eq:ansatz}
P_{\pm}(x,t) = P_{\pm}(x|v_0(t),k_0,D/\alpha(t))^*,
\end{equation}
where $k_0 = k(t=0)$. In particular, if we choose boundary conditions such that 
\[\alpha(t=0) = \alpha(t=\tau) = 1,\]
and
\[k(t=0) = k(t=\tau) = k_0,\]
this ansatz allows for a rapid transition between steady states corresponding to different velocities. We highlight that this ansatz is \textit{not} in steady-state for intermediate times, i.e. for $0<t<\tau$, however, it \textit{is} at $t=0$ and $t=\tau$ with the boundary conditions specified. We now evaluate Eq.~\eqref{eq:master_eq} for this ansatz. The algebra is somewhat lengthy and we relegate the full calculation to the supplementary material, however, the final necessary constraints on the control parameters are relatively simple:
\begin{subequations}
\label{eq:shortcut}
\begin{align}
& \dot{\alpha} = 2\alpha \mu (k -k_0\alpha), \\
&\dot{v}_0 =-v_0\mu ( k-k_0).
\end{align}
\end{subequations}
Remarkably, this set of control is \textit{identical} to that found in \cite{baldovin_2022} for 2D ABPs. ABPs and RTPs are known to exhibit equivalent hydrodynamics and share many physical properties at a large scales \cite{Cates_EPL_2013,Solon_EPJ_2015}, though we emphasize that this shortcut is identical at the single-particle, noninteracting level for different dynamics, equilibrium distributions, and perhaps most surprisingly, dimensionality. This suggests that Eq.~\eqref{eq:shortcut} may prove to be a universal shortcut for both of these systems, or an even broader class of active systems, irrespective of specific model details.

\begin{figure*}
    \centering
    \includegraphics[width=\linewidth]{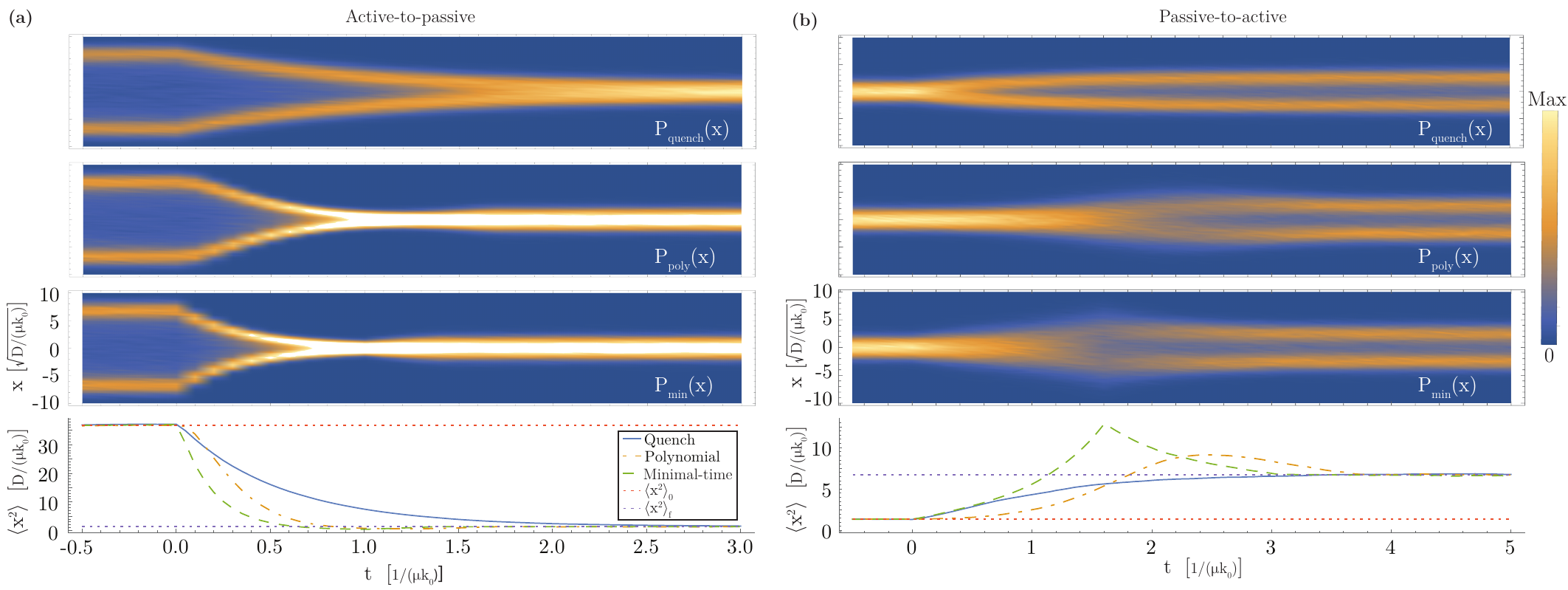}
    \caption{Shortcut protocols can outperform quenched protocols in Langevin dynamics simulations. (Left) Particle distribution as a function of time from transition onset ($t=0$) shown for an active-to-passive transition with $\mu k_0 =1/2$, $\gamma = 1/5$, $d=1$, $v_{00} = 5$ and $v_{0f} = 1/2$ in simulation units. Rapid parameter quench and resulting relaxation (top),  polynomial shortcut with $\tau_{\text{poly}} = 1.7 1/(\mu k_0)$ (second from top), and minimal-time shortcut with the same minimum and maximum stiffness as the polynomial protocol with $\tau_{\text{min}} \approx 1.31 1/(\mu k_0)$ (third from top). The polynomial protocol was chosen such that $k(t)>0$ at all times. Bottom subplot shows the position variance for each of the simulated protocols, quench in solid blue, polynomial in dashed-dotted orange, and minimal-time in dashed green, as well as the theoretical initial, dotted orange, and final, dotted purple, variances (see \cite{SM} for derivation). Note that both polynomial and minimal-time shortcut protocols converge to the final target distribution significantly faster than quenched relaxation. (Right) Same as left plots but for a passive-to-active transition with $v_{00} = 1/2$ and $v_{0f} = 2$, and we find $\tau_{\text{poly}} = 3.85 1/(\mu k_0)$ and $\tau_{\text{min}} = 3.04 1/(\mu k_0)$, again chosen so $k_{\text{poly}}(t)>0$. Note that for this transition, the minimal-time shortcut protocol marginally outperforms a naive parameter quench, as shown by the slightly more rapid convergence of the dashed green curve relative to the solid blue curve to the final target value in the bottom subplot. 
    Moreover, the polynomial shortcut \textit{underperforms} the naive quench for this transition. Therefore, the active-to-passive shortcut significantly outperforms the natural quenched relaxation, but the passive-to-active shortcut presents a more complicated picture.}
    \label{fig:fig2}
\end{figure*}

\indent \textit{Explicit protocols. }--
With Eq.~\eqref{eq:shortcut} in hand, we are now able to solve for explicit shortcut protocols for this system. Moreover, given the identical solution to the ABP case, we may map to the solutions already presented \cite{baldovin_2022} We first consider a polynomial solution for $\alpha$ to Eq.~\eqref{eq:shortcut}. Considering the four boundary conditions that must be satisfied in addition to the first order differential equation, we require five degrees of freedom and therefore minimally a four degree polynomial (though higher degree inequivalent solutions will likewise produce viable shortcuts). Evaluating Eq.~\eqref{eq:shortcut} for this polymonial ansatz yields
\begin{equation}
\label{eq:polynomial}
\alpha(t) = 1+\frac{30t^2(t-\tau)^2\log \frac{v_{00}}{v_{0f}}}{\mu k_0 \tau^5},
\end{equation}
where $v_{00} (v_{0f})$ is the initial (final) velocity of the protocol, and $\tau$ is the protocol duration. Given $\alpha$, one can immediately find $k(t)$ and $v_0(t)$ by manipulating Eq.~\eqref{eq:shortcut}. However, we note one potential issue with this solution: for $v_{0f}>v_{00}$, that is when the final velocity is greater than the initial, $\alpha(t)$ can, in principle, become zero or even go negative, which, in turn, leads to singularities in $k(t)$ and $v_0(t)$. Evaluating Eq.~\eqref{eq:polynomial}, the extrema occur at $t=0,\tau/2,$ and $\tau$. For $v_{0f}>v_{00}$, the minimum of this function occurs at $\tau/2$ which yields a constraint on the minimum protocol duration for such a solution:
\begin{equation}
\alpha(t)>0 \implies \tau > \frac{15}{8}\frac{\log \frac{v_{00}}{v_{0f}}}{\mu k_0}.
\end{equation}
We remark that the typical timescale for relaxation to equilibrium is given by $\tau_r \sim 1/(\mu k_0)$ such that a polynomial shortcut may not significantly outperform, or may even be slower than, a simple quenched protocol.

In contrast, when $v_{0f}<v_{00}$, there is no such constraint on the minimal time, and, in principle, arbitrarily fast transitions are possible. Such transitions, however, may require physically inaccessible protocols for $k(t)$, such as very strong or even negative stiffness values, but there are no singularities that prevent
their mathematical construction.

Given a constrained minimum and maximum value for stiffness, denoted $k_{\text{min}}$ and $k_{\text{max}}$, respectively, a ``bang-bang" optimal minimal time protocol was previously constructed in \cite{baldovin_2022}. In particular, in the case of $v_{0f}>v_{00}$ $(v_{0f}<v_{00})$ the protocol consists of spending a certain amount of time $t_1$ at $k_\text{min}$ ($k_{\text{max}}$) before instantaneously transitioning to $k_\text{max}$ ($k_{\text{min}}$) for the remainder $\tau_*-t_1$ of the protocol. Again one finds $\tau_*$ is larger for $v_{0f}>v_{00}$ than for $v_{0f}<v_{00}$, even for the same values of $k_{\text{max}}$ and $k_{\text{min}}$.

In Fig.~\ref{fig:fig2}, we have plotted simulations of Langevin dynamics for both active-to-passive (left), $v_{0f}<v_{00}$, and passive-to-active (right), $v_{0f}>v_{00}$ transitions for a quenched transition (top), polynomial shortcut transition (middle), and minimal-time shortcut transition (bottom) for identical maximum and minimum stiffness values as the polynomial protocol. Observe that for passive-to-active transitions, the shortcut protocols are not significantly faster than simple quenched relaxation, though there is a significant speedup for active-to-passive transitions, in principle only limited by the external control.

\begin{figure}
    \centering
    \includegraphics[width=\columnwidth]{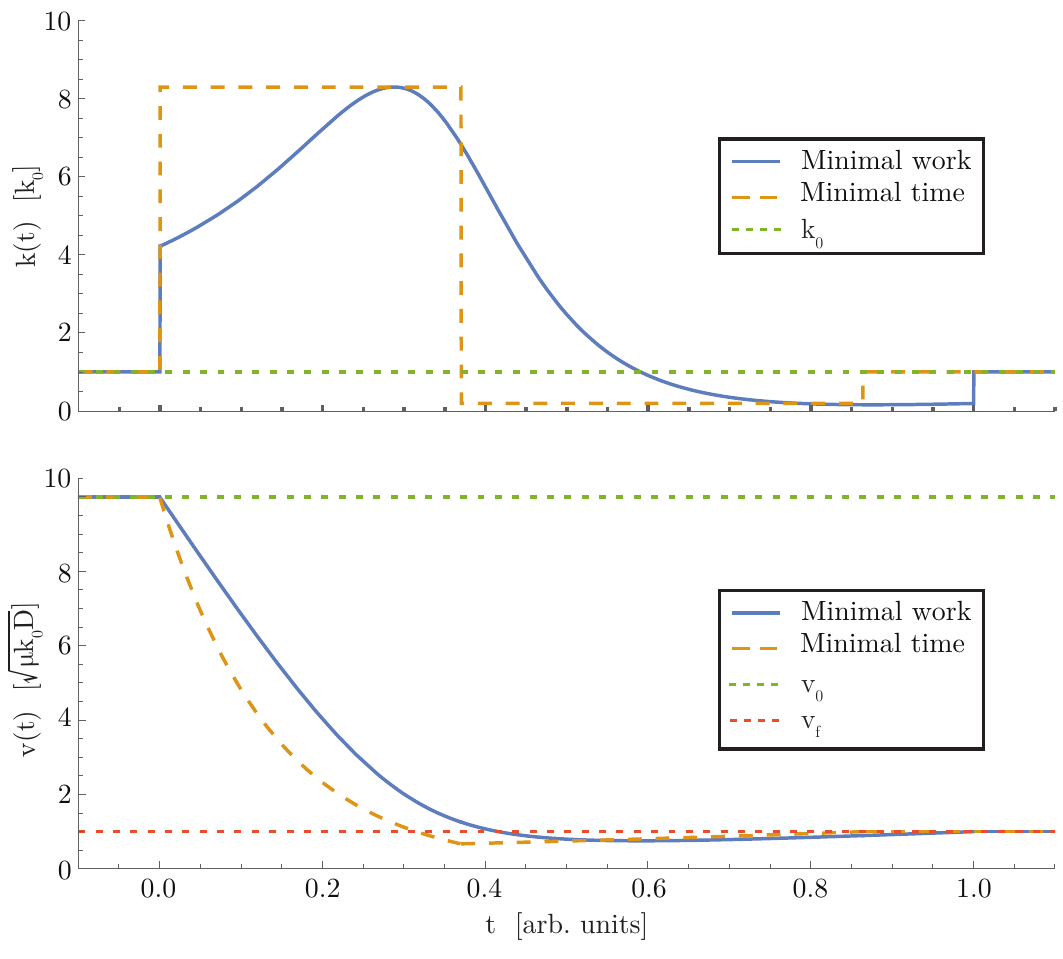}
    \caption{Minimal work (solid blue) and Minimal time (dashed orange) protocols for $k(t)$ and $v_0(t)$. Maximum and minimum allowed values of stiffness are chosen to be identical and time is measured with respect to duration of minimal-work protocol. We find $\tau_{\text{min-time}} \approx 0.864 \tau_{\text{min-work}}$ while $W_\text{min-time} \approx 1.571  W_{\text{min-work}}$.}
    \label{fig:fig3}
\end{figure}

\indent \textit{Minimal work protocol.}--
We now consider the energetic cost of such shortcuts. The external work of a given shortcut protocol is given by (see \cite{SM} for a full derivation),
\begin{multline}
\label{eq:work}
\braket{W} =  \int_0^{\tau}  \int_{-\infty}^\infty dx (\partial_t U(x) P(x,t)) \\
=\int_0^{\tau} \frac{\dot{k}}{2}\left[\frac{D}{\alpha \mu k _0} + \frac{v_0^2}{(\mu k_0)^2+ \mu k_0 \gamma }\right]dt ,\\
\end{multline}
where $v_0(t),k(t)$ and $\alpha(t)$ still must satisfy Eq.~\eqref{eq:shortcut}. Defining
\begin{equation}
    \Lambda(t) \equiv \int_0^{t}\alpha(t')dt',
\end{equation}
we can write $\braket{W}$ as a local functional:
\begin{widetext}
\begin{equation}
\label{eq:functional}
\braket{W} = \int_0^{\tau} \left(
\frac{\dddot{\Lambda}}{2\mu\dot{\Lambda}}-\frac{\ddot{\Lambda}^2}{2\mu\dot{\Lambda}^2} +k_0\ddot{\Lambda}
\right)\left[\frac{D}{2\mu k_0\dot{\Lambda}} + \frac{v_{00}^2}{(\mu k_0)^2+\mu k_0 \gamma)}\frac{1}{\dot{\Lambda}}e^{2\mu k_0 t}e^{-2\Lambda} \right]dt,
\end{equation}
\end{widetext}
from which one may derive Euler-Lagrange equations to minimize the external work. We relegate the full expression to the supplemental material, but it consists of a nonlinear fourth-order ordinary differential equation in $\Lambda(t)$. Specifying the boundary conditions $\Lambda(0) = 0, \dot{\Lambda}(0) = \dot{\Lambda}(\tau) = 1$, and $\Lambda(\tau) = \tau-1/(\mu k_0)\log(v_{0f}/v_{00})$ ensures the system begins in the steady state $P(x|v_{00},k_0,D)$ and ends in $P(x|v_{0f},k_0,D)$ in total duration $\tau$ with minimal external work. We show the required protocols for $k(t)$ and $v_0(t)$ for such a minimal-work protocol in Fig.~\ref{fig:fig3}and compare to a minimal-time protocol with identical values of $k_{\text{min}}$ and $k_{\text{max}}$. Observe the discontinuities in $k$ at the start and end of protocol, which have previously been observed in finite-time optimal control problems \cite{2007_PRL_Schmiedl,Blabler_PRE_2021,Zhong_PRE_2022}. Further, this protocol contrasts significantly with the minimal-time protocol, indicating a trade-off between protocol duration and work, a ubiquitous feature of finite-time thermodynamics \cite{2016_PRL_Proesmans,2018_PRL_Pietzonka,Shiraishi_PRL_2018}.

\indent \textit{Discussion.}--
By applying the ansantz $P(x,t) = P(x|v_0(t),k_0,D/\alpha(t))$, where $\alpha(0) = \alpha(\tau) = 1$ and $k(0) = k(\tau) = k_0$, we have derived a class of shortcut solutions, Eq.~\eqref{eq:shortcut}, that allow for rapid transitions between 1D RTP steady states at the same trap stiffness and temperature and different self-propulsion velocities. Although the limit to $D\rightarrow 0$ (athermal limit) is nontrivial, we emphasize that the same shortcut solution works in this case with $\alpha \rightarrow 0$ (see \cite{SM} for derivation). In future work, it will be interesting to consider similar questions for changing trap stiffness or temperature. As a start, although we focused on protocols with $D$ held fixed, one could instead quench $D$ at the end of the protocol to some $D' = D/\alpha(\tau)$, which allows for transitions between different-temperature steady states. We leave the study of shortcut protocols with smoothly-varying temperature for future work.

It is striking that Ref. \cite{baldovin_2022} derived the \textit{identical} shortcut solution as Eq.~\eqref{eq:shortcut} for 2D ABPs in harmonic confinement. RTPs and ABPs are known to share hydrodynamic properties at large length scales, though we emphasize that our work and Ref. \cite{baldovin_2022} focused on \textit{distinct} single-particle distributions in \textit{different} dimensions: it is surprising to us that the shortcut protocols turn out to be identical for these distinct systems. This may point to a universal shortcut protocol applicable to numerous active systems, which could prove useful in engineering active matter. We emphasize that there are very few fully calculated steady-state distributions for active particle systems, such that the typical inverse engineering techniques in shortcut methods will need to be augmented to study general active matter shortcuts.

Another observation worthy of comment is the tendency of passive-to-active shortcuts to take more time than passive-to-active shortcuts. In fact, if we allow any value for harmonic trap stiffness for intermediate times, including unphysical negative values which have been previously studied in the literature \cite{2018_NJP_Chupeau}, we find that there is no lower time bound whatsoever on active-to-passive transitions. We suggest a simple physical intuition for this asymmetry: in passive-like steady states, both right- and left-moving particles are concentrated at the center of the trap. In contrast, in active-like steady states, right- (left-) moving particles accumulate at the right (left) wall of the trap. Therefore, when trying to transition from an active state to a passive state, one can simply increase the stiffness dramatically to accumulate all particles at the center of the trap before decreasing slightly and allowing the particles to reach their proper steady states. Indeed, we obtained this qualitative strategy quite generally for all explicit protocols we calculated, as shown in Fig~\ref{fig:fig3}. In contrast, when transitioning from a passive state to an active state, one can only decrease the stiffness to $k = 0$ (at a minimum) and allow the two particle species to reach their final positions by means of their self-propulsion before again raising the stiffness to maintain the trap. This is qualitatively the general protocol we found for optimal passive to active transitions. We emphasize that even if we were to allow $k$ to attain negative values, this may prove detrimental to rapid passive-to-active transitions as a left-moving particles could become initially trapped at the right wall, and vice versa. Conversely, negative values of stiffness allow for arbitrarily rapid active-to-passive transitions.

In this work, we calculated the minimal-time and minimal-work protocols when constrained to our shortcut solution Eq.~\eqref{eq:ansatz}. More specifically, we focused on the minimal external work protocol: we did not take into account the energy dissipated from the active particles themselves or from other external sources, such as external controls needed to modulate self-propulsion velocities. Including such energetic costs may lead to different optimal protocols. We emphasize that this shortcut, Eq.~\eqref{eq:shortcut}, offers a great deal of flexibility given the arbitrary possible choice of $\alpha$. That said, although this family of solutions does yield full, rapid control of the distribution, it may not contain the absolutely optimal set of controls for a given state-to-state transition. In particular, we calculated the minimal-time and minimal-work protocols connecting two different RTP steady states for a given duration $\tau$ when constrained to this specific control strategy. We were able to observe presumably general features, such as work and time trade-offs, but it is a worthwhile question to discover and engineer absolutely optimal protocols for controlling active matter, perhaps through approaches from optimal transport theory \cite{Aurell_PRL_2011,Zhong_PRE_2022}. For example, the observed asymmetry in passive-to-active and active-to-passive transitions may be a feature of this specific control protocol, but may not prove to be truly universal. Furthermore, there exists a large literature on optimal control of passive matter in the slowly-varying regime where systems deviate only slightly from reference equilibrium states \cite{2012_PRL_Sivak,2012_PRE_zulkowski,2013_PLOS_Zulkowski,2015_PRE_Rotskoff,2020_PRL_Brandner,2020_abiuso_prl,Frim_PRE_2022,Frim_PRL_2022}. Extending these results to active systems could mark a significant advance in engineering active systems.\\

\begin{acknowledgments}
\textit{Acknowledgements.} 
AGF is supported by the NSF GRFP under Grant No. DGE 1752814. This work was supported in part by the U.S. Army Research Laboratory and the U.S. Army Research Office under contract W911NF-20-1-0151.

\end{acknowledgments}

\end{document}


\preprint{APS/123-QED}

\title{Supplemental Material for ``Shortcut engineering of active matter: run-and-tumble particles"}

\author{Adam G. Frim}%
\affiliation{%
 Department of Physics, University of California, Berkeley, Berkeley, California, 94720
}%
\author{Michael R. DeWeese}
\affiliation{%
 Department of Physics, University of California, Berkeley, Berkeley, California, 94720
}%
\affiliation{%
Redwood Center For Theoretical Neuroscience,  University of California, Berkeley, Berkeley, California, 94720
}
\affiliation{%
Helen Wills Neuroscience Institute, University of California, Berkeley, Berkeley, California, 94720
}%

\date{\today}

\maketitle

\section{Derivation}
We start with the master equation coupling the right- and left-moving particles due to particle transmutation, Eq.~(3) of the main text,
\begin{align}
\label{eq:FK_1}
&\partial_t P_+ = -\partial_x[(-\mu k x+v_0)P_+] + D\partial_x^2 P_+ -\frac{\gamma}{2}(P_+-P_-) \\
\label{eq:FK_2}
&\partial_t P_- = -\partial_x[(-\mu k x-v_0)P_-] + D\partial_x^2 P_- -\frac{\gamma}{2}(P_--P_+) .
\end{align}
This has the stationary state solution given by \cite{Garcia-Millan_JSM_2021}

\begin{equation}
    P_{\pm}(x)^* = P_{\pm}^*(x|v_0,k,D) = \sum_{n=0}^\infty\lambda^n \mathcal{P}_{\pm}^{(n)} =  \sum_{n=0}^\infty \left(\frac{v_0}{\sqrt{ \mu k _0 D}} \right)^n \frac{1}{2} \sqrt{\frac{\mu k }{2\pi D}}e^{-\frac{\mu k  x^2}{2D}}  (\pm 1)^n C_n(\frac{\gamma}{\mu k })  \textit{He}_n\left(\sqrt{\frac{\mu k }{D}} x\right),
\end{equation}
where $C_0(x) \equiv 1$ and
\begin{equation}
    C_n(x) = 
    \prod_{j=1}^n \frac{1}{j+\frac{1}{2}(1+(-1)^{n+1})x}, \ \ \ \  n>0.
\end{equation}
$\textit{He}_n(x)$ is the $n$th Hermite polynomial under the ``probabilist's convention,"
\[\textit{He}_n(x) = (-1)^n e^{\frac{x^2}{2}} \frac{d^n}{dx^n}e^{-\frac{x^2}{2}}.\]
We define $\lambda \equiv v_0/\sqrt{\mu k_0 D}$ and 
\[\mathcal{P}_\pm^{(n)} \equiv \frac{1}{2} \sqrt{\frac{\mu k }{2\pi D}}e^{-\frac{\mu k  x^2}{2D}}  (\pm 1)^n C_n(\frac{\gamma}{\mu k })  \textit{He}_n\left(\sqrt{\frac{\mu k }{D}} x\right). \]
Observe that the coefficients $C_n$ satisfy the recurrence relation 
\begin{equation}
C_n(x) = 
    \frac{1}{n+\frac{1}{2}(1+(-1)^{n+1})x} C_{n-1}(x),
\end{equation}
for $n>0$. We may further define $C_n=0$ for $n<0$. \\

\noindent
To verify our solution, we evaluate this differential equation for our ansatz:
\begin{equation}
    P_{\pm}(x,t) = P_{\pm}^*(x|v_0(t), k _0,D/\alpha(t)) = \sum_{n=0}^\infty \left(\frac{\sqrt{\alpha(t)} v_0(t)}{\sqrt{\mu k _0 D}} \right)^n \frac{1}{2} \sqrt{\frac{\alpha(t) \mu k _0}{2\pi D}}e^{-\frac{\alpha(t)\mu k _0x^2}{2D}}  (\pm 1)^n C_n(\frac{\gamma}{\mu k _0})  \textit{He}_n\left(\sqrt{\frac{\mu k _0\alpha(t)}{D}} x\right).
\end{equation}
To aid in our derivation, let us calculate several derivatives that will be helpful later. For clarity, we suppress the arguments of $\textit{He}_n$ as they will all be of the same form, $\sqrt{\mu k_0 \alpha/D} x$:
\begin{equation}
    \partial_t P_{\pm} = \sum_{n=0}^\infty \lambda^n \left[(n+1)\frac{\dot{\alpha}}{2\alpha}+ \frac{n\dot{v}_0 }{v_0}- \frac{\mu k _0x^2}{2D}\dot{\alpha} + nx\sqrt{\frac{\alpha \mu k _0}{D}} \frac{ \textit{He}_{n-1}}{\textit{He}_n}\frac{\dot{\alpha}}{2\alpha} \right] \mathcal{P}^{(n)},
\end{equation}
\begin{equation}
    \partial_x P_{\pm} = \sum_{n=0}^\infty \lambda^n \left[-\frac{\alpha \mu k _0 x}{D}+ n\sqrt{\frac{\alpha \mu k _0}{D}}\frac{ \textit{He}_{n-1}}{\textit{He}_n} \right] \mathcal{P}_\pm^{(n)},
\end{equation}
\begin{equation}
    \partial_x^2 P_{\pm} = \sum_{n=0}^\infty \lambda^n \left[-\frac{\alpha \mu k _0}{D} + \left(\frac{\alpha \mu k _0 }{D}\right)^2 x^2- 2n x\left(\frac{\alpha \mu k _0}{D} \right)^{3/2} \frac{\textit{He}_{n-1}}{\textit{He}_n} + n(n-1) \frac{\alpha \mu k _0}{D} \frac{\textit{He}_{n-2}}{\textit{He}_n} \right] \mathcal{P}_\pm^{(n)}.
\end{equation}
Using the recursion identity 
\[\textit{He}_{n+1}(y) = y \textit{He}_n(y) - n \textit{He}_{n-1}(y) \implies \textit{He}_{n-2}(y) = \frac{y \textit{He}_{n-1}(y) - \textit{He}_n(y)}{n-1}\]
we may simplify further
\begin{equation}
\begin{split}
    \partial_x^2 P_{\pm}& \sum_{n=0}^\infty \lambda^n \left[-\frac{\alpha \mu k _0}{D} + \left(\frac{\alpha \mu k _0 }{D}\right)^2 x^2- 2n x\left(\frac{\alpha \mu k _0}{D} \right)^{3/2} \frac{\textit{He}_{n-1}}{\textit{He}_n} + n(n-1) \frac{\alpha \mu k _0}{D} \frac{\textit{He}_{n-2}}{\textit{He}_n} \right] \mathcal{P}_\pm^{(n)}\\
    &= \sum_{n=0}^\infty \lambda^n \left[-\frac{\alpha \mu k _0}{D} + \left(\frac{\alpha \mu k _0 }{D}\right)^2 x^2- 2n x\left(\frac{\alpha \mu k _0}{D} \right)^{3/2} \frac{\textit{He}_{n-1}}{\textit{He}_n} + n x\left(\frac{\alpha \mu k _0}{D}\right)^{3/2} \frac{\textit{He}_{n-1}}{\textit{He}_n}  - \frac{n\alpha \mu k _0}{D} \right] \mathcal{P}_\pm^{(n)}\\
    & = \sum_{n=0}^\infty \lambda^n \left[-(n+1)\frac{\alpha \mu k _0}{D} + \left(\frac{\alpha \mu k _0 }{D}\right)^2 x^2- n x\left(\frac{\alpha \mu k _0}{D} \right)^{3/2} \frac{\textit{He}_{n-1}}{\textit{He}_n}  \right] \mathcal{P}_\pm^{(n)}.\\
    \end{split}
\end{equation}
Finally, we calculate
\begin{equation}
\begin{split}
v_0 P_{\pm} = \sum_{n=0}^\infty (\sqrt{\frac{\mu k _0 D}{\alpha}} \lambda )\lambda^n \mathcal{P}_{\pm}^{(n)} = \sum_{n=0}^\infty \lambda^{n+1} \sqrt{\mu k _0 D} \mathcal{P}_\pm^{(n)} 
&= 
\sum_{n=0}^\infty \lambda^{n+1} \sqrt{\frac{\mu k _0 D}{\alpha}} \frac{\mathcal{P}_\pm^{(n)}}{\mathcal{P}_\pm^{(n+1)}}\mathcal{P}_\pm^{(n+1)}\\
& = \sum_{n=1}^\infty \lambda^{n} (\pm 1) \sqrt{\frac{\mu k _0 D}{\alpha}} \frac{C_{n-1}}{C_{n}} \frac{\textit{He}_{n-1}}{\textit{He}_n}\mathcal{P}_\pm^{(n)}\\
&= \sum_{n=0}^\infty \lambda^{n} (\pm 1) \sqrt{\frac{\mu k _0 D}{\alpha}} \frac{C_{n-1}}{C_{n}} \frac{\textit{He}_{n-1}}{\textit{He}_n}\mathcal{P}_\pm^{(n)}\\
&= \sum_{n=0}^\infty \lambda^{n} (\pm 1) \sqrt{\frac{\mu k _0 D}{\alpha}} (n+\frac{\gamma}{2\mu k _0}(1+(-1)^{n+1})) \frac{\textit{He}_{n-1}}{\textit{He}_n}\mathcal{P}_\pm^{(n)},\\
\end{split}
\end{equation}
such that
\begin{equation}
\begin{split}
\pm \partial_x (v_0 P_{\pm}) 
& = \sum_{n=0}^\infty \lambda^{n}  \sqrt{\frac{\mu k _0 D}{\alpha}} (n+\frac{\gamma}{2\mu k _0}(1+(-1)^{n+1})) \left[\frac{-\alpha \mu k _0 x}{D} \frac{\textit{He}_{n-1}}{\textit{He}_n}+(n-1)\sqrt{\frac{\alpha \mu k _0}{D}}\frac{\textit{He}_{n-2}}{\textit{He}_n} \right] \mathcal{P}_\pm^{(n)}\\
&= \sum_{n=0}^\infty \lambda^{n} \sqrt{\frac{\mu k _0 D}{\alpha}} (n+\frac{\gamma}{2\mu k _0}(1+(-1)^{n+1})) \left(-\sqrt{\frac{\alpha \mu k _0}{D}}\right) \mathcal{P}_\pm^{(n)}\\
&= -\sum_{n=0}^\infty \lambda^{n}  (n \mu k _0+\frac{\gamma}{2}(1+(-1)^{n+1}))  \mathcal{P}_\pm^{(n)},\\
\end{split}
\end{equation}
where we have made use of the recurrence relation.

Putting these all together then, we find, to satisfy the master equation:
\begin{equation}
\begin{split}
&
 \begin{split}
 0 = \sum_{n=0}^\infty 
&\bigg[
\left((n+1)\frac{\dot{\alpha}}{2\alpha}+ \frac{n\dot{v}_0 }{v_0} - \frac{\mu k _0x^2}{2D}\dot{\alpha} + nx\sqrt{\frac{\alpha \mu k _0}{D}}\frac{ \textit{He}_{n-1}}{\textit{He}_n}\frac{\dot{\alpha}}{2\alpha}\right) -\mu k \\
& \ \ \  -\mu k x\left(-\frac{\alpha \mu k _0 x}{D}+ n\sqrt{\frac{\alpha \mu k _0}{D}}\frac{ \textit{He}_{n-1}}{\textit{He}_n}\right)- (n \mu k _0+\frac{\gamma}{2}(1+(-1)^{n+1}))\\
& \ \ \ -D\left(-(n+1)\frac{\alpha \mu k _0}{D} + \left(\frac{\alpha \mu k _0 }{D}\right)^2 x^2- n x\left(\frac{\alpha \mu k _0}{D} \right)^{3/2} \frac{\textit{He}_{n-1}}{\textit{He}_n}\right)+\frac{\gamma}{2}(1+(-1)^{n+1})
\bigg] \lambda^n \mathcal{P}_{\pm}
\end{split}
\\
&\begin{split}
& \phantom{0 } = 
\sum_{n=0}^\infty 
\bigg[\left(\frac{\dot{\alpha}}{2\alpha} +\alpha \mu k _0-\mu k  \right) \left(-\frac{\alpha \mu k _0x^2}{D} + nx \sqrt{\frac{\alpha \mu k _0}{D}}\frac{\textit{He}_{n-1}}{\textit{He}_n}+1\right) + n\left(\frac{\dot{\alpha}}{2\alpha}+\frac{\dot{v}_0}{v_0}-\mu k _0+\alpha \mu k _0 \right)
\bigg]\lambda^n \mathcal{P}_{\pm}
\end{split}
\end{split}
\end{equation}
We may satisfy this equation for 
\begin{align}
& \frac{\dot{\alpha}}{2\alpha} +\alpha \mu k _0-\mu k  = 0, \\
& \frac{\dot{\alpha}}{2\alpha}+\frac{\dot{v}_0}{v_0}-\mu k _0+\alpha \mu k _0 = 0.
\end{align}
Note that by solving the first, we can further simplify the second:
\begin{align}
\label{eq:S_shortcut_1}
& \dot{\alpha} = 2\mu \alpha ( k  - k _0\alpha), \\
\label{eq:S_shortcut_2}
&\dot{v}_0 = - v_0 \mu (  k - k _0).
\end{align}
Remarkably, this is precisely the same set of conditions needed for the active Brownian particle shortcut \cite{baldovin_2022}.

\section{Athermal Shortcut}

For the athermal case, to avoid nuances in taking the limit of $D\rightarrow 0$, we can instead work directly with the athermal distribution  \cite{Dhar_PRE_2019}:

\begin{equation}
P_{\pm}(x)^* = P_{\pm}^*(x|v_0,k) = \frac{1}{2} \frac{\mu k \Gamma[\frac{1}{2}(1+\frac{\gamma}{\mu k})]}{\sqrt{\pi} v_0 \Gamma[\frac{\gamma}{2\mu k}]}
\left(1\pm \frac{\mu k x}{v_0} \right)^{\frac{2\gamma}{\mu k}}\left(1\mp \frac{\mu k x}{v_0} \right)^{\frac{2\gamma}{\mu k}-1}.
\end{equation}
We simply use as an ansatz
\begin{equation}
P_\pm(x,t) = P_\pm^*(x|v_0(t),k_0) =  \frac{1}{2} \frac{\mu k \Gamma[\frac{1}{2}(1+\frac{\gamma}{\mu k})]}{\sqrt{\pi} v_0 \Gamma[\frac{\gamma}{2\mu k}]} \left(1\pm \frac{\mu k_0 x}{v_0(t)} \right)^{\frac{2\gamma}{\mu k_0}}\left(1\mp \frac{\mu k_0 x}{v_0(t)} \right)^{\frac{2\gamma}{\mu k_0}-1}.
\end{equation}
Evaluating these ans\"{a}tze in Eqs.~\eqref{eq:FK_1}~and~\eqref{eq:FK_2}, for $D\rightarrow 0$ we have, after simplification,
\begin{align}
& \frac{\mu k_0   \Gamma \left[\frac{1}{2} \left(\frac{\gamma }{\mu k_0 }+1\right)\right] \left(\dot{v}_0 +v_0 \mu (k-k_0)  \right) \left(1-\frac{\mu^2 k_0^2 x^2}{v_0^2}\right)^{\frac{\gamma }{2 \mu k_0  }-1} \left(\gamma  \mu k_0   x^2- \mu  k_0 V_0 x-v_0^2)\right)}{v_0^3 \Gamma \left[\frac{\gamma }{2 k_0 \mu }\right]}=0,\\
 & \frac{\mu k_0   \Gamma \left[\frac{1}{2} \left(\frac{\gamma }{\mu k_0 }+1\right)\right] \left(\dot{v}_0 +v_0\mu (k-k_0)  \right) \left(1-\frac{\mu^2 k_0^2 x^2}{v_0^2}\right)^{\frac{\gamma }{2 \mu k_0  }-1} \left(\gamma  \mu k_0   x^2+ \mu  k_0 V_0 x-v_0^2\right)}{v_0^3 \Gamma \left[\frac{\gamma }{2 k_0 \mu }\right]}=0.
\end{align}
Both equations may be naturally satisfied by taking

\begin{equation}
    \dot{v}_0 = -v_0 \mu (k-k_0),
\end{equation}
which is precisely Eq.~\eqref{eq:S_shortcut_2}, or the shortcut upon taking $\alpha \rightarrow 0$. 

\section{Average external work}
We now calculate the average work done externally in this protocol. We emphasize that we focus only on the work due to manipulating the potential: there may also be some energy cost in changing $v_0$ (not to mention also the housekeeping heat to maintain the system's activity), but herein we focus only on the potential work. Therefore,
\begin{multline}
\braket{W} = \int_0^{t_f} dt \ \int dx \ P(x,t) \partial_t U(x,t) \\= \int_0^{t_f} \ \frac{\dot{k}}{2} \int dx \  x^2   \sum_{n=0}^\infty \left(\frac{\sqrt{\alpha(t)} v_0(t)}{\sqrt{\mu k _0 D}} \right)^{2n}  \sqrt{\frac{\alpha(t) \mu k _0}{2\pi D}}e^{-\frac{\alpha(t)\mu k _0x^2}{2D}}  C_{2n}(\frac{\gamma}{\mu k _0})  \textit{He}_n\left(\sqrt{\frac{\mu k _0\alpha(t)}{D}} x\right).
\end{multline}
Let us focus on the spatial integral
\[\begin{split}
\int dx \  x^2e^{-\frac{\alpha(t)\mu k _0 x^2}{2D}} \textit{He}_{2n}\left(\sqrt{\frac{\mu k _0\alpha(t)}{D}} x\right)& = \left(\frac{D}{\mu k }\right)^{3/2} \int dy \  y^2e^{-y^2/2} \textit{He}_{2n}(y)\\
&= \left(\frac{D}{\alpha(t) \mu k _0}\right)^{3/2} \int dy \  y^2\frac{d^{2n}}{dy^{2n}} e^{-y^2/2} = \sqrt{2\pi}\left(\frac{D}{\alpha(t)\mu k _0}\right)^{3/2}(\delta_{n,0}+2\delta_{n,1}).
\end{split}\]
Accordingly, the work performed during the protocol is 
\begin{equation}
\begin{split}
\braket{W} &=  \int_0^{t_f} dt \ \frac{\dot{k}}{2} \sum_{n=0}^\infty \left(\frac{\sqrt{\alpha(t)} v_0(t)}{\sqrt{\mu k _0 D}} \right)^{2n}  \sqrt{\frac{\alpha(t) \mu k _0}{2\pi D}} C_{2n}(\frac{\gamma}{\mu k _0}) \sqrt{2\pi}\left(\frac{D}{\alpha(t)\mu k _0}\right)^{3/2}(\delta_{n,0}+2\delta_{n,1})\\
& = \int_0^{t_f} \ \frac{\dot{ k }}{2}\left[\frac{D}{\alpha \mu k _0} + \frac{v_0^2}{(\mu k_0)^2+ \mu k_0 \gamma }\right]dt. \\
\end{split}
\end{equation}
From Eqs.~\eqref{eq:S_shortcut_1}~and~\eqref{eq:S_shortcut_2}, we have
\[v_0(t) = v_{00} e^{-\mu \int_0^{t}(k-k_0)dt},\]
where $v_{00}$ is the initial velocity. We can also rearrange:
\[k = \frac{\dot{\alpha}}{2\mu \alpha} + k_0\alpha \implies \dot{k} = \frac{\ddot{\alpha}}{2\mu\alpha} - \frac{\dot{\alpha}^2}{2\mu\alpha^2}+k_0\dot{\alpha},\]
such that
\[v_0(t) = v_{00} e^{- \int_0^{t} [\frac{\dot{\alpha}}{2\alpha}+\mu k_0(\alpha(t')-1)]dt'} = \frac{v_{00}}{\sqrt{\alpha}}e^{\mu k_0 t} e^{-\mu k_0 \int_0^t \alpha(t')dt'}.\]
Finally, defining
\begin{equation}
\Lambda(t) = \int_0^t \alpha(t') dt',
\end{equation}
we can rewrite $\braket{W}$ as a local functional in $\Lambda$:
\begin{equation}
\braket{W} = \int_0^{t_f} \left(
\frac{\dddot{\Lambda}}{2\mu\dot{\Lambda}}-\frac{\ddot{\Lambda}^2}{2\mu\dot{\Lambda}^2} +k_0\ddot{\Lambda}
\right)\left[\frac{D}{2\mu k_0\dot{\Lambda}} + \frac{v_{00}^2}{(\mu k_0)^2+\mu k_0 \gamma)}\frac{1}{\dot{\Lambda}}e^{2\mu k_0 t}e^{-2\Lambda} \right]dt,
\end{equation}
such that we may derive the Euler-Lagrange equation for $\Lambda(t)$:

\begin{equation}
    \begin{split}
       0= &D \left(6 \ddot{\Lambda}^3+\ddddot{\Lambda} \dot{\Lambda}^2-6 \dddot{\Lambda} \dot{\Lambda} \ddot{\Lambda}\right)+\frac{2 v_{00}^2 e^{2 \mu k_0 (s-\Lambda)}}{\gamma +\mu k_0} \left(8 (\mu k_0)^3 \dot{\Lambda}^6-20 (\mu k_0)^3 \dot{\Lambda}^5+16 (\mu k_0)^3 \dot{\Lambda}^4-4 \mu k_0 \dot{\Lambda}^3 \left((\mu k_0)^2+3 \mu k_0 \ddot{\Lambda}+\dddot{\Lambda}\right)\right)\\
       &
       +\frac{2 v_{00}^2 e^{2 \mu k_0 (s-\Lambda)}}{\gamma +\mu k_0} \left(\dot{\Lambda}^2 \left(8 (\mu k_0)^2 \ddot{\Lambda}+4 \mu k_0 \dddot{\Lambda}+6 \mu k_0 \ddot{\Lambda}^2+\ddddot{\Lambda}\right)-3 \dot{\Lambda} \ddot{\Lambda} \left(3 \mu k_0 \ddot{\Lambda}+2 \dddot{\Lambda}\right)+6 \ddot{\Lambda}^3\right).
    \end{split}
\end{equation}
Solving this Euler-Lagrange equation will lead to extremal work protocols, such as the minimal protocol constructed in the main text.